\documentclass[runningheads,a4paper]{llncs}
\usepackage{makeidx}
\usepackage{amssymb}
\usepackage{amscd, amsfonts}
\usepackage{verbatim}
\usepackage{listings}
\usepackage{commath}
\usepackage{float}
\usepackage{tikz}

\usepackage{lmodern}
\usepackage[T1]{fontenc}

\newcommand{\primen}{\text{prime }}
\newcommand{\dvd}{\text{ dvd }}
\renewcommand{\set}{\operatorname{set}}

\allowdisplaybreaks[1]

\begin{document}

\title{$\textrm{AUTO2}$, a saturation-based heuristic prover for
  higher-order logic}

\author{Bohua Zhan}

\institute{Massachusetts Institute of Technology}
\maketitle

\begin{abstract}
  We introduce a new theorem prover for classical higher-order logic
  named \texttt{auto2}. The prover is designed to make use of
  human-specified heuristics when searching for proofs. The core
  algorithm is a best-first search through the space of propositions
  derivable from the initial assumptions, where new propositions are
  added by user-defined functions called proof steps. We implemented
  the prover in Isabelle/HOL, and applied it to several formalization
  projects in mathematics and computer science, demonstrating the high
  level of automation it can provide in a variety of possible proof
  tasks.
\end{abstract}

\section{Introduction}

The use of automation is a very important part of interactive theorem
proving. As the theories to be formalized become deeper and more
complex, having a good automatic tool becomes increasingly
indispensable. Such tools free users from the tedious task of
specifying low level arguments, allowing them to focus instead on the
high level outline of the proof.

There is a large variety of existing automatic proof tools. We will be
content to list some of the representative ones. Some tools emulate
human reasoning by attempting, at any stage of the proof, to apply a
move that humans are also likely to make. These include the
\texttt{grind} tactic in PVS \cite{grind}, and the ``waterfall''
algorithm in ACL2 \cite{ACL2}. A large class of automatic provers are
classical first-order logic solvers, based on methods such as tableau,
satisfiability-modulo-theories (SMT), and superposition calculus.
Sledgehammer in Isabelle \cite{hammering} is a representative example
of the integration of such solvers into proof assistants. Finally,
most native tools in Isabelle and Coq are based on tactics, and their
compositions to realize a search procedure. Examples for these include
the \texttt{auto} tactic in Isabelle and Coq. The \texttt{blast}
tactic in Isabelle \cite{blast} can also be placed in this category,
although it has some characteristics of classical first-order solvers.

All these automatic tools have greatly improved the experience of
formalization using proof assistants. However, it is clear that much
work still needs to be done. Ideally, formalizing a proof on the
computer should be very much like writing a proof in a textbook, with
automatic provers taking the place of human readers in filling in any
``routine'' intermediate steps that are left out in the proof. Hence,
one reasonable goal for the near future would be to develop an
automatic prover that is strongly enough to consistently fill in such
intermediate steps.

In this paper, we describe an alternative approach toward automation
in proof assistants. It is designed to combine various desirable
features of existing approaches. On the one hand it is able to work
with human-like heuristics, classical higher-order logic, and simple
type theory. On the other hand it has a robost, saturation-based
search mechanism. We discuss these features and their motivations in
Section \ref{sec:objectives}.

As a first approximation, the algorithm in our approach consists of a
best-first search through the space of propositions derivable from the
initial assumptions, looking for a contradiction (any task is first
converted into contradiction form). New propositions are generated by
\emph{proof steps}: user provided functions that match one or two
existing propositions, and produce new propositions that logically
follow from the matched ones. The order in which new propositions are
added is dictated by a scoring function, as in a best-first search
framework. There are several elaborations to this basic picture, in
order to support case analysis, rewriting, skolemization, and
induction. The algorithm will be described in detail, along with a
simple example, in Section \ref{sec:description}.

We implemented our approach in Isabelle/HOL, and used it to develop
several theories in mathematics and computer science. In these case
studies, we aim to use \texttt{auto2} to prove all major theorems,
either on its own or using a proof outline at a level of detail
comparable to that of human exposition. We believe this aim is largely
achieved in all the case studies. As a result, the level of automation
provided by \texttt{auto2} in our examples compares favorably with,
and in some cases greatly exceeds that of existing tools provided in
Isabelle. We give some examples from the case studies in Section
\ref{sec:examples}.

The implementation, as well as the case studies, are available at
\[\texttt{https://github.com/bzhan/auto2}.\] We choose the name \texttt{auto2}
for two reasons: first, we intend it to be a general purpose prover
capable of serving as the main automatic tool of a system, as
\texttt{auto} in Isabelle and Coq had been. Second, it relates to one
of the main features of the algorithm, which is that any proof step
matches at most two items in the state.

In Section \ref{sec:relatedwork}, we compare our approach with other
major approaches toward automation, as well as list some related
work. We conclude in Section \ref{sec:conclusion}, and discuss
possible improvements and future directions of research.

\section{Objectives}
\label{sec:objectives}

In this section, we list the main features our approach is designed to
have, and the motivations behind these features.

\paragraph{Use of human-like heuristics:} The prover should make use
of heuristics that humans employ when searching for proofs. Roughly
speaking, such heuristics come in two levels. At the lower level,
there are heuristics about when to apply a single theorem. For
example, a theorem of the form $A\implies B\implies C$ can be applied
in three ways: deriving $C$ from $A$ and $B$, deriving $\neg A$ from
$B$ and $\neg C$, and deriving $\neg B$ from $A$ and $\neg C$. Some of
these directions may be more fruitful than others, and humans often
instinctively apply the theorem in some of the directions but not in
others. At the higher level, there are heuristics concerning
induction, algebraic manipulations, procedures for solving certain
problems, and so on. Both levels of heuristics are essential for
humans to work with any sufficiently deep theory. Hence we believe it
is important for the automatic prover to be able to take these into
account.

\paragraph{Extensibility:} The system should be extensible in the
sense that users can easily add new heuristics. At the same time, such
additions should not jeopardize the soundness of the prover. This can
be guaranteed by making sure that every step taken by the user-added
heuristics is verified, following the LCF framework.

\paragraph{Use of higher-order logic and types:} The prover should be
able to work with higher-order logic, and any type information (in the
Isabelle sense) that is present. In particular, we want to avoid
translations to and from untyped first-order logic that are
characteristic of the use of classical first-order solvers. Avoiding
these has several benefits: many heuristics that humans use are best
stated in higher-order logic. Also, the statement to be proved is kept
short and close to what humans work with, which facilitates printing
an informative trace when a proof fails.

\paragraph{Saturation-based search mechanism:} Most heuristics are
fallible in the sense that they are not appropriate in every
situation, and can lead to dead ends when applied in the wrong
situations. Moreover, when several mutually-exclusive heuristics are
applicable, we would like to consider all of them in turn. Some kind
of search is necessary to deal with both of these problems. We follow
a saturation-based search strategy in order to obtain the following
desirable property: all steps taken by the prover are both permanent
and ``non-committal''. That is, the result of any step is available
for use throughout the remainder of the search, but there is never a
requirement for it to be used, to allow for the possibility that the
step is not appropriate for the proof at hand. The choice of
E-matching over simplification to deal with equality reasoning is also
chosen with this property in mind.

\paragraph{~~~} Having listed the principles motivating our approach,
we also want to clarify what are not our main concerns. First, our
focus is on proof tasks that occur naturally as intermediate steps
during proofs of theorems in mathematics and computer science. We do
not intend the prover to be competitive against more specialized
algorithms when faced with large tasks that would also be difficult
for humans.  Second, the prover is not fully automated in the sense
that it requires no human intervention -- the user still needs to
provide heuristics to the prover, including how to use each of the
previously proved theorems. Finally, we do not intend to make the
prover complete. For more difficult theorems, it expects hints in the
form of intermediate steps.

\section{Description of the system}
\label{sec:description}

In this section, we describe the \texttt{auto2} prover in detail,
followed by a simple example, and a discussion of how the system is
used in practice. We begin with a high-level description of the
algorithm, leaving the details to the following subsections.

The algorithm follows a saturation-based strategy, maintaining and
successively adding to a list of \emph{items}. We will call this list
the \emph{main list} in the remainder of this section. For a first
pass, we can think of items as propositions that follow from the
initial assumptions, and possibly additional assumptions. Later on
(Section \ref{sec:itemtypes}) we will see that it can also contain
other kinds of information, in addition to or instead of a
proposition. Each item is placed in a \emph{box}, which specifies what
additional assumptions the item depends on. We discuss boxes in more
detail in Section \ref{sec:boxlattice}.

New items that may be added to the list are created by \emph{proof
  steps}, which are user-provided functions that accept as input one
or two existing items, and derive a list of new items from the
inputs. With a few exceptions (Section \ref{sec:exists_inst}), the new
items must logically follow from the input items. One common kind of
proof steps matches the input items to the one or two assumptions of a
theorem, and when there is a match, return the conclusion of the
theorem. However, as proof steps are arbitrary functions, they can
have more complex behavior.

Reasoning with equalities is achieved by matching up to equivalence
(E-matching) using a \emph{rewrite table}. The rewrite table is a data
structure that maintains the list of currently known equalities (not
containing schematic variables). It provides a matching function that,
given a pattern $p$ and a term $t$, returns all matches of $t$ against
$p$, up to rewriting $t$ using the known equalities. The rewrite table
automatically uses transitivity of equality, as well as the congruence
property (that is, $a_1=b_1,\dots,a_n=b_n$ implies
$f(a_1,\dots,a_n)=f(b_1,\dots,b_n)$). See \cite{ematching} for a
modern introduction to E-matching. In our implementation, E-matching
is essentially a first-order process (we only make use of equalities
between terms not in function position), but we also allow matching of
certain higher-order patterns, and extend it in other ways (Section
\ref{sec:matching}). Matching using the rewrite table is used as the
first step of nearly all proof steps.

New items produced by proof steps are collected into \emph{updates},
and each update is assigned a score, which indicates its priority in
the best-first search. All new updates are first inserted into a
priority queue. At each iteration of the algorithm, the update with
the lowest score is pulled from the queue. The items contained in the
update are then added to the main list and processed
one-by-one. Scoring is discussed in Section \ref{sec:scoring}.

With these in mind, we can give a first sketch of the main loop of the
algorithm. We assume that the statement to be proved is written in
contradiction form (that is, $[A_1,\dots,A_n]\implies C$ is written as
$[A_1,\dots,A_n,\neg C]\implies\texttt{False}$), so the goal is to
derive a contradiction from a list of assumptions $A_1,\dots,A_n$.

\begin{itemize}
\item The algorithm begins by inserting a single update to the
  priority queue, containing the propositions $A_1,\dots, A_n$.

\item At each iteration, the update with the lowest score is pulled
  from the priority queue. Items within the update are added
  one-by-one to the main list.

\item Upon adding a non-equality item, all proof steps taking one
  input item are invoked on the item. All proof steps taking two input
  items are invoked on all pairs of items consisting of the new item
  and another item in the main list. All updates produced are added to
  the priority queue.

\item Upon adding an equality item (without schematic variables), the
  equality is added to the rewrite table. Then the procedure in the
  previous step is redone with the new rewrite table on all items
  containing up to equivalence either side of the equality (this is
  called \emph{incremental matching}). All new updates (those that
  depend on the new equality) are added to the priority queue.

\item The loop continues until a contradiction (depending only on the
  initial assumptions) are derived by some proof step, or if there are
  no more updates in the queue, or if some timeout condition is
  reached.

\end{itemize}

In the current implementation, we use the following timeout condition:
the loop stops after pulling $N$ updates from the priority queue,
where $N$ is set to 2000 (in particular, all invocations of
\texttt{auto2} in the given examples involve less than 2000 steps).

\subsection{Box lattice}\label{sec:boxlattice}

Boxes are used to keep track of what assumptions each item depends
on. Each \emph{primitive} or \emph{composite} box represents a list of
assumptions. They are defined recursively as follows: a composite box
is a set of primitive boxes, representing the union of their
assumptions. The primitive boxes are indexed by integers starting at
0. Each primitive box inherits from a composite box consisting of
primitive boxes with smaller index, and contains an additional list of
assumptions. It represents the result of adding those assumptions to
the parent box. The primitive box 0 (inheriting from \{\}) contains
the list of assumptions in the statement to be proved. Other primitive
boxes usually inherit, directly or indirectly, from \{0\}. The
primitive boxes also keep track of introduced variables. From now on
we will simply call a composite box as a \emph{box}.

If a contradiction is derived in a box (that is, if \texttt{False} is
derived from the assumptions in that box), the box is called
\emph{resolved}, and appropriate propositions (negations of the
assumptions) are added to each of its immediate parent boxes. The
overall goal of the search is then to resolve the box \{0\}, which
contains exactly the assumptions for the statement to be proved.

There is a natural partial order on the boxes given by inclusion, and
a merge operation given by taking unions, making the set of boxes into
a semilattice. New primitive boxes are created by proof steps, and are
packaged into updates and added to the queue with a score just like
new items. Creating a new primitive box effectively starts a case
analysis, as we will explain in the example in Section
\ref{sec:simpleexample}.

\subsection{Item types} \label{sec:itemtypes}

In this section we clarify what information may be contained in an
item. In general, we think of an item in a box $b$ as any kind of
information that is available under the assumptions in $b$. One
important class of items that are not propositions are the term
items. A term item $t$ in box $b$ means $t$ appears as a subterm of
some proposition (or another kind of item) in $b$. The term items can
be matched by proof steps just like propositions. This allows the
following implementation of directed rewrite rules: given a theorem $P
= Q$, where any schematic variable appearing in $Q$ also appears in
$P$, we can add a proof step that matches $P$ against any term item
$t$, and produces the equality $P(\sigma) = Q(\sigma)$ for any match
with instantiation $\sigma$. This realizes the forward rewrite rule
from $P$ to $Q$.

In general, each item consists of the following information: a string
called \emph{item type} that specifies how to interpret the item; a
term called \emph{tname} that specifies the content of the item; a
theorem that justifies the item if necessary, and an integer score
which specifies its priority in the best-first search. The most basic
item type is \texttt{PROP} for propositions, for which \emph{tname} is
the statement of the theorem, and is justified by the theorem
itself. Another basic type is \texttt{TERM} for terms items, for which
\emph{tname} is the term itself, and requires no justifying theorems.

The additional information contained in items can affect the behavior
of proof steps, and by outputting an item with additional information,
a proof step can affect how the output is used in the future. This
makes it possible to realize higher level controls necessary to
implement more complex heuristics. To give a simple example, in the
current implementation, disjunctions are stored under two different
item types: \texttt{DISJ} and \texttt{DISJ\_ACTIVE}. The latter type
induces case analysis on the disjunction, while the former does
not. By outputting disjunctions in the appropriate type, a proof step
can control whether case analysis will be invoked on the result.

\subsection{Skolemization and induction}
\label{sec:exists_inst}

Usually, when a proof step outputs a proposition, it must derive the
justifying theorem for that proposition, using the justifying theorems
of the input items. There are two main exceptions to this. First,
given an input proposition $\exists x.\, P(x)$, a proof step can
output the proposition $P(x)$, where $x$ is a previously unused
constant. This realizes skolemization, which in our framework is just
one of the proof steps.

The second example concerns the use of certain induction theorems. For
example, induction on natural numbers can be written as:
\[ P(0) \implies \forall n.\, P (n - 1) \longrightarrow P(n) \implies
P(n).
\] This form of the induction theorem suggests the following method of
application: suppose $n$ is an initial variable in a primitive box
$i$, and proposition $n\neq 0$ is known in (the composite) box
$\{i\}$. Then we may insert $P(n-1)$ into box $\{i\}$, where $P$ is
obtained from the list of assumptions in $i$ containing $n$. This
corresponds to the intuition that once the zero case is proved, one
may assume $P(n-1)$ while proving $P(n)$.

In both cases, any contradiction that depends on the new proposition
can be transformed into one that does not. In this first case, this
involves applying a particular theorem about existence (\texttt{exE}
in Isabelle). In the second case, it involves applying the induction
theorem.

\subsection{Matching} \label{sec:matching}

In this section, we provide more details about the matching
process. First, the presence of box information introduces additional
complexities to E-matching. In the rewrite table, each equality is
stored under a box, and each match is associated to a box, indicating
which assumptions are necessary for that match. When new items are
produced by a proof step, the items are placed in the box that is the
merge of boxes containing the input items, and the boxes associated to
all matches performed by that proof step.

We also support the following additional features in matching:
\begin{itemize}
\item Matching of associative-commutative (AC) functions: the matching
  makes limited use of properties of AC functions. For example, if $x
  = y \star z$ is known, where $\cdot\star\cdot$ is AC, then the
  pattern $y \star ?a$ can match the term $p \star x$, with
  instantiation $?a := p \star z$ (since $y \star (p \star z) = p
  \star (y \star z) = p \star x$). The exact policy used in
  AC-matching is rather involved, as it needs to balance efficiency
  and not missing important matches.

\item Matching of higher-order patterns: we support second-order
  matching, with the following restriction on patterns: it is possible
  to traverse the pattern in such a way that any schematic variable in
  function position is applied to distinct bound variables in its
  first appearance. For example, in the following theorem:
  \[ \forall (n::\text{nat}).\, f(n) \le f (n + 1) \implies m \le n
  \implies f(m) \le f(n), \]

  one can match its first assumption and conclusion against two items,
  since the left side of the inequality in the first assumption can be
  matched to give a unique instantiation for $f$. The condition given
  here is slightly more general than the condition given by Nipkow
  \cite{pattern}, where all appearances of a schematic variable in
  function position must be applied to distinct bound variables.

\item Schematic variables for numeric constants: one can restrict a
  schematic variable to match only to numeric constants (in the
  current implementation, this is achieved by a special name
  $?\mathtt{NUMC}_i$). For example, one can write proof steps that
  perform arithmetic operations, by matching terms to patterns such as
  $?\mathtt{NUMC}_1 + ?\mathtt{NUMC}_2$.

\item Custom matching functions: one can write custom functions for
  matching a pattern against an item. This is especially important for
  items of type other than \texttt{PROP}. But it is also useful for
  the \texttt{PROP}s themselves. For example, if the pattern is
  $\neg(p < q)$, one can choose to match $q \le p$ instead, and
  convert any resulting theorem using the equivalence to $\neg(p <
  q)$.
\end{itemize}

\subsection{Scoring} \label{sec:scoring}

The scoring function, which ranks future updates, is crucial for the
efficiency of the algorithm as it determines which updates will be
explored first in the search. It tries to guess which reasoning steps
are more likely to be relevant to the proof at hand. In the current
implementation, we choose a very simple strategy. Finding a better
scoring strategy will certainly be a major focus in the future.

The current scoring strategy is as follows: the score of any update
equals the maximum of the scores of the dependent items, plus an
increment depending on the content of the update. The increment is
bigger (i.e. the update is discouraged) if the terms in the update are
longer, or if the update depends on many additional assumptions.

\subsection{A simple example} \label{sec:simpleexample}

We now give a sample run of \texttt{auto2} on a simple theorem. Note
this example is for illustration only. The actual implementation
contains different proof steps, especially for handling
disjunctions. Moreover, we ignore scoring and the priority queue,
instead adding items directly to the list. We also ignore items that
do not contribute to the eventual proof.

The statement to be proved is
\[ \primen p \implies p > 2 \implies \text{odd } p. \] Converting to
contradiction form (and noting that odd $p$ is an abbreviation for
$\neg \text{even } p$), our task is to derive a contradiction from
assumptions $\primen p$, $p > 2$, and $\text{even } p$. The steps are:

\begin{enumerate}
\item Add primitive box 0, with variable $p$, and assumptions $\primen
  p$, $p > 2$, and $\text{even } p$.

\item Add subterms of the propositions, including $\texttt{TERM }
  \primen p$ and $\texttt{TERM } \text{even } p$.

\item The proof step for applying the definition of prime adds
  equality \[\primen p = (p > 1 \wedge \forall m.\, m \dvd p
  \longrightarrow m = 1 \vee m = p)\] from $\texttt{TERM } \primen
  p$. Likewise, the proof step for applying the definition of even
  adds equality $\text{even } p = 2 \dvd p$ from $\texttt{TERM }
  \text{even } p$.

\item When the first equality in the previous step is applied,
  incremental matching is performed on the proposition $\primen p$. It
  now matches the pattern $?A \wedge ?B$, so the proof step for
  splitting conjunctions produces $p > 1$ and $\forall m.\, m \dvd p
  \longrightarrow m = 1 \vee m = p$.

\item A proof step matches the propositions $\forall m.\, m \dvd p
  \longrightarrow m = 1 \vee m = p$ and $\text{even } p$ (the second
  item, when rewritten as $2 \dvd p$, matches the antecedent of the
  implication), producing $2 = 1 \vee 2 = p$.

\item \label{item:figurestep} The proof step for invoking case
  analysis matches $2 = 1 \vee 2 = p$ with pattern $?A \vee ?B$. It
  creates primitive box 1, with assumption $2 = 1$ (see Figure
  \ref{fig:simpleexample}).

\begin{figure}[H]
\centering
\begin{tikzpicture}
  \node[draw,align=center,anchor=south west] at (0,0) {
    \textbf{Box \{0\}}\\
    $\primen p$\\
    $p > 2$\\
    $\text{even } p$\\
    \texttt{TERM } $\primen p$\\
    \texttt{TERM } $\text{even } p$\\
    $\primen p = p > 1 \wedge \dots$ \\
    $\text{even } p = 2 \dvd p$ \\
    $\forall m.\, m \dvd p \longrightarrow m = 1 \vee m = p$\\
    $2 = 1 \vee 2 = p$ };

  \node[draw,align=center,anchor=south west] at (7,0) {
    \textbf{Box \{1\}}\\
    $2 = 1$ };

  \draw [->] (6.9,0.5) -- (5,0.5);

  \node[align=left,anchor=north west] at (5.5,4) {
    Primitive box 0: \\
    Variables: $p$ \\
    Assumptions: $\primen p$, $p > 2$, $\text{even } p$ \\ \\
    Primitive box 1 (inherit from \{0\}): \\
    Variables: --\\
    Assumptions: $2 = 1$};
\end{tikzpicture}
\caption{State of proof after step \ref{item:figurestep}. Arrow
  indicates inheritance relation on boxes.}
\label{fig:simpleexample}
\end{figure}
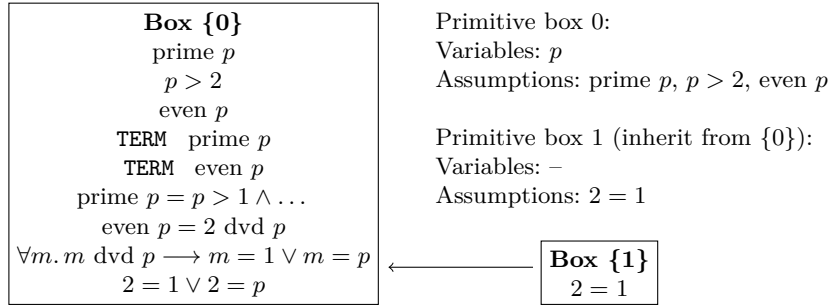

\item A proof step matches $2 = 1$ (in box $\{1\}$) with pattern
  $?\mathtt{NUMC}_1 =\: ?\mathtt{NUMC}_2$. The proof step examines the
  constants on the two sides, finds they are not equal, and outputs a
  contradiction. This resolves box $\{1\}$, adding $2\neq 1$ into box
  $\{0\}$.

\item A proof step matches $2 = 1 \vee 2 = p$ with $2\neq 1$,
  producing $2 = p$.

\item When the equality in the previous step is added, incremental
  matching is performed on the proposition $p > 2$ (one of the initial
  assumptions). This proposition matches pattern $?n >\: ?n$ (when
  rewritten as $p > p$ or $2 > 2$), giving a contradiction. This
  resolves box $\{0\}$ and finishes the proof.
\end{enumerate}

\subsection{Proof scripts}

For the case studies, we designed our own language of proof scripts
for specifying intermediate steps in the proof of a more difficult
theorem. The proof scripts are provided as an argument to the
\texttt{auto2} tactic, and are interpreted within the tactic. This
requires some straightforward modifications to the main loop and the
scoring mechanism, which we will not discuss. The benefit of using an
internally interpreted script (instead of Isar) is that the entire
state of the proof is maintained between lines of the script, and all
previously proved statements are available for use at any given point.

The proof script consists of atomic commands joined together with two
connectors: \texttt{THEN} and \texttt{WITH}. Each atomic command
specifies an intermediate statement to prove, and what update to add
once that statement is proved. The meanings of the two connectors are
as follows. The command $A$ \texttt{THEN} $B$ means first process $A$,
and after $A$ is finished, process $B$. The command $A$ \texttt{WITH}
$B$ (with $A$ atomic) means attempt to prove the intermediate
statement specified in $A$, processing $B$ as a part of the attempt.

The simplest atomic commands are \texttt{OBTAIN} and
\texttt{CASE}. The command \texttt{OBTAIN} $p$ means attempt to prove
$p$ and add it to the list. The command \texttt{CASE} $p$ means
attempt to prove that $p$ results in a contradiction, and add $\neg p$
to the list. It is equivalent to \texttt{OBTAIN} $\neg p$.

The command \texttt{CHOOSE} $x,\, p(x)$ specifies $\exists x.\, p(x)$ as an
intermediate statement. After it is proved, the resulting existence
fact is instantiated with variable $x$ (the command fixes variable $x$
so it is not used in other places).

Finally, there are various flavors of induction commands, which
specify applications of various kinds of induction theorems. We
designed the script system to be extensible: it is possible for the
user to add new types of atomic commands.

\subsection{Practical usage} \label{sec:usage}

We end this section with a discussion of practical issues concerning
the use of the \texttt{auto2} system.

First, we describe the process of constructing the collection of proof
steps. The collection of proof steps specifies exactly what steps of
reasoning \texttt{auto2} may take. With the exception of equality
reasoning, which relies on the rewrite table and E-matching, all other
forms of reasoning are encoded as proof steps. This includes basic
deductions in logic and arithmetic, and the simplification of
terms. In particular, \texttt{auto2} does not invoke any of the other
Isabelle commands such as \texttt{simp} and \texttt{arith}, except
within the implementation of individual proof steps, for carrying out
very specific tasks.

Each proof step is intended to represent a single step of reasoning,
and has a clearly-defined behavior. The simplest proof steps apply a
single theorem. For example, a theorem of the form $A\implies
B\implies C$ can be added for use in either the forward or one of the
two backward directions. More complex proof steps are implemented as
ML functions. The implementation can make full use of the existing
conversion and tactics facility in Isabelle/ML.

In theories developed using \texttt{auto2}, each proof step using
theorems in that theory is added right after all required theorems are
proved. Once the proof step is added, it is used in all ensuing
proofs, both in the current theory and in all descendent theories. For
theorems proved in the Isabelle library, ``wrapper'' theories are
created to add proof steps using them. The case studies, for example,
use shared wrapper theories for theorems concerning logic, arithmetic,
sets, and lists.

There are some circumstances where removing a proof step after using
it in a few proofs is acceptable. For example, if a theory introduces
constructions, or proves lemmas that are used only within the theory,
it is acceptable to remove proof steps related to those constructions
and lemmas once they are no longer used. The guiding principle is as
follows: by the end of the development of a theory, the collection of
proof steps from that theory should form a coherent system of
heuristics on how to use the results in that theory. In subsequent
theories, \texttt{auto2} should have a basic competence in using
results from that theory, and it should always be possible to specify
more involved applications in proof scripts. In particular, the user
should never need to add proof steps for using theorems from a
previous theory, nor temporarily remove a proof step from a previous
theory (to avoid exploding the search space). Realizing this principle
means more work is needed when building each theory, to specify the
right set of proof steps, but it should pay off in the long run, as it
frees the user from having to refer back to the theory in subsequent
developments.

Second, we describe the usual interaction loop when proving a theorem
for which applying \texttt{auto2} directly fails. One begins by
working out an informal proof of the theorem, listing those steps that
appear to require some creativity. One can then try \texttt{auto2}
with these intermediate steps added. If it still does not work, the
output trace shows the first intermediate step that \texttt{auto2}
cannot prove, and what steps of reasoning are taken in the attempt to
prove that step. If there is some step of reasoning that should be
taken automatically but is not, it is an indication that some proof
step is missing. The missing proof step should be added, either to a
wrapper theory if the relevant theorem is proved in the Isabelle
library, or right after the theorem if it is proved in a theory
developed using \texttt{auto2}. On the other hand, if one feels the
missing step should not be taken automatically, but is a non-obvious
step to take in the proof of the current theorem, one should add that
step to the proof script instead. The process of adding to the
collection of proof steps or to the proof script continues until
\texttt{auto2} succeeds.

\section{Case studies}\label{sec:examples}

In this section, we give some examples from the case studies conducted
using \texttt{auto2}. We will cover two of the six case
studies. Descriptions for the other four (functional data structures,
Hoare logic, construction of real numbers, and Arrow's impossibility
theorem) can be found in the repository. In writing the case studies,
we aim to achieve the following goal: all major theorems are proved
using \texttt{auto2}, either directly or using proof scripts at a
level of detail comparable to human exposition. When a case study
parallels an existing Isabelle theory, there may be some differences
in the definitions, organization, and method of proof used. The
content of the theorems, however, are essentially the same. In the
examples below, we will sometimes compare the length of our scripts
with the length of Isar scripts for the same theorem in the Isabelle
library. We emphasize that this is not intended to be a rigorous
comparison, due to the differences just mentioned, and since
\texttt{auto2} is provided additional information in the form of the
set of proof steps, and takes longer to verify the script. The intent
is rather to demonstrate the level of automation that can be expected
from \texttt{auto2}.

Besides the examples given below, we also make a special note of the
case study on Arrow's impossibility theorem. The corresponding theory
in the Isabelle AFP is one of the seven test theories used in a series
of benchmarks on Sledgehammer, starting in \cite{judgement}.

\subsection{Elementary theory of prime numbers}

The development of the elementary theory of prime numbers is one of
the favourites for testing theorem provers \cite{boyer,mizar}. We
developed this theory starting from the definition of prime numbers,
up to the proof of the infinitude of primes and the unique
factorization theorem, following \texttt{HOL/Number\_Theory} in the
Isabelle library. For the infinitude of primes, the main lemma is that
there always exists a larger prime:
\begin{lstlisting}
larger_prime: $\exists p.\,\primen p \wedge n < p$
\end{lstlisting}

\texttt{auto2} is able to prove this theorem when provided with the
following proof script:
\begin{lstlisting}
CHOOSE $p,\, \primen p \wedge p \dvd \text{fact } n + 1$ THEN
CASE $p \le n$ WITH OBTAIN $p \dvd \text{fact } n$
\end{lstlisting}

This corresponds to the following proof of \texttt{next\_prime\_bound}
in the Isabelle theory \texttt{HOL/Number\_Theory/Primes} (18 lines).

\begin{lstlisting}
lemma next_prime_bound: $\exists p.\, \primen p \wedge n < p \wedge p \le \text{fact } n + 1$
proof-
  have f1: $\text{fact } n + 1 \neq$ (1::nat)" using fact_ge_1 [of n, where 'a=nat] by arith
  from prime_factor_nat [OF f1]
  obtain $p$ where $\primen p$ and $p \dvd \text{fact } n + 1$ by auto
  then have $p \le \text{fact } n + 1$ apply (intro dvd_imp_le) apply auto done
  { assume $p \le n$
    from $\primen p$ have $p \ge 1$
      by (cases $p$, simp_all)
    with $p \le n$ have $p \dvd \text{fact } n$
      by (intro dvd_fact)
    with $p \dvd \text{fact } n + 1$ have $p \dvd \text{fact } n + 1 - \text{fact } n$
      by (rule dvd_diff_nat)
    then have $p \dvd 1$ by simp
    then have $p \le 1$ by auto
    moreover from $\primen p$ have $p > 1$
      using prime_def by blast
    ultimately have False by auto}
  then have $n < p$ by presburger
  with $\primen p$ and $p \le \text{fact } n + 1$ show ?thesis by auto
qed
\end{lstlisting}

Likewise, we formalized the unique factorization theorem. The
uniqueness part of the theorem is as follows (note $M$ and $N$ are
multisets, and $\set M$ and $\set N$ are the sets corresponding to $M$
and $N$, eliminating duplicates).
\begin{lstlisting}
factorization_unique_aux:
$\forall p\in \set M.\, \primen p \implies
\forall p\in \set N.\, \primen p \implies \prod_{i\in M} i \dvd
\prod_{i\in N} i \implies M \subseteq N$
\end{lstlisting}

The script needed for the proof is:
\begin{lstlisting}
CASE $M = \emptyset$ THEN
CHOOSE $M',\, m, \,M = M' + \{m\}$ THEN
OBTAIN $m \dvd \prod\nolimits_{i\in N} i$ THEN
CHOOSE $n,\, n\in N \wedge m \dvd n$ THEN
CHOOSE $N',\, N = N' + \{n\}$ THEN
OBTAIN $m = n$ THEN
OBTAIN $\prod\nolimits_{i\in M'} i \dvd \prod\nolimits_{i\in N'} i$ THEN
STRONG_INDUCT $(M, [\text{Arbitrary } N])$
\end{lstlisting}

This can be compared to the proof of
\texttt{multiset\_prime\_factorization\_unique\_aux} in the Isabelle
theory \texttt{HOL/Number\_Theory/UniqueFactorization} (39 lines).

\subsection{Verification of imperative programs}

A much larger project is the verification of imperative programs,
building on the Imperative HOL library, which describes imperative
programs involving pointers using a Heap Monad \cite{imphol}. The
algorithms and data structures verified are:

\begin{itemize}
\item Reverse and quicksort on arrays.
\item Reverse, insert, delete, and merge on linked lists.
\item Insert and delete on binary search trees.
\end{itemize}

The proofs are mostly automatic, which is in sharp contrast with the
corresponding examples in the Isabelle distribution (in
\texttt{Imperative\_HOL/ex}). We give one example here. The merge
function on two linked lists is defined as:

\begin{lstlisting}
partial_function (heap) merge :: ('a::{heap, ord}) node ref $\Rightarrow$ 'a node ref $\Rightarrow$ 'a node ref Heap where
[code]: merge p q =
  do { np $\leftarrow$ !p; nq $\leftarrow$ !q;
       if np = Empty then return q
       else if nq = Empty then return p
       else if val np $\le$ val nq then
         do { npq $\leftarrow$ merge (nxt np) q;
              p := Node (val np) npq;
              return p }
       else
         do { pnq $\leftarrow$ merge p (nxt nq);
              q := Node (val nq) pnq;
              return q } }
\end{lstlisting}

To prove the main properties of the merge function, we used the
following two lemmas (commands adding their proof steps are omitted):

\begin{lstlisting}
theorem set_intersection_list: (x $\cup$ xs) $\cap$ ys = {} $\Rightarrow$ xs $\cap$ ys = {} by auto

theorem unchanged_outer_union_ref:
  "unchanged_outer h h' (refs_of h p $\cup$ refs_of h q) $\Rightarrow$ r $\notin$ refs_of h p $\Rightarrow$ r $\notin$ refs_of h q $\Rightarrow$
   Ref.present h r $\Rightarrow$ Ref.get h r = Ref.get h' r" by (simp add: unchanged_outer_ref)
\end{lstlisting}

The statements of the theorems are:
\begin{lstlisting}
theorem merge_unchanged:
  "effect (merge p q) h h' r $\Rightarrow$ proper_ref h p $\Rightarrow$ proper_ref h q $\Rightarrow$
   unchanged_outer h h' (refs_of h p $\cup$ refs_of h q)"

theorem merge_local:
  "effect (merge p q) h h' r $\Rightarrow$ proper_ref h p $\Rightarrow$ proper_ref h q $\Rightarrow$
   refs_of h p $\cap$ refs_of h q = {} $\Rightarrow$ proper_ref h' r $\wedge$ refs_of h' r $\subseteq$ refs_of h p $\cup$ refs_of h q"

theorem merge_correct:
  "effect (merge p q) h h' r $\Rightarrow$ proper_ref h p $\Rightarrow$ proper_ref h q $\Rightarrow$
   refs_of h p $\cap$ refs_of h q = {} $\Rightarrow$ list_of h' r = merge_list (list_of h p) (list_of h q)"
\end{lstlisting}

Each of these theorems is proved (in 30--40 seconds on a laptop) using
the same proof script, specifying the induction scheme:
\begin{lstlisting}
DOUBLE_INDUCT (("pl = list_of h p", "ql = list_of h q"), Arbitraries ["p", "q", "h'", "r"])
\end{lstlisting}

In the Isabelle library the proof of the three corresponding theorems,
including that of two induction lemmas proved specifically for these
theorems, takes 166 lines in total. These theorems also appear to be
well beyond the ability of the Sledgehammer tools. It is important to
note that this automation is not based on Hoare logic or separation
logic (the development here is separate from the case study on Hoare
logic), but the proofs here use directly the semantics of commands
like in the original examples.

\section{Related work}\label{sec:relatedwork}

The author is particularly inspired by the work of Ganesalingam and
Gowers \cite{gowers}, which describes a theorem prover that can output
proofs in a form extremely similar to human exposition. Our
terminology of ``box'' is taken from there (although the meaning here
is slightly different).

There are two ways in which our approach resembles some of the
classical first-order solvers. The first is the use of a
``blackboard'' maintaining a list of propositions, with many
``modules'' acting on them, as in a Nelson-Oppen architecture
\cite{smt}. The second is the use of matching up to equivalence
(E-matching), which forms a basic part of most SMT solvers. The main
differences are explained in the first three items in Section
\ref{sec:objectives}: our focus on the use of human-like heuristics,
and our lack of translation to and from untyped first-order logic.

There have been extensive studies on heuristics that humans use when
proving theorems, and their applications to automation. Ganesalingam
and Gowers \cite{gowers} give a nice overview of the history of such
efforts. Some of the more recent approaches include the concept of
proof plans introduced by Alan Bundy \cite{Bundy1},
\cite{Bundy2}. Among proof tools implemented in major proof
assistants, the \texttt{grind} tactic \cite{grind} and the
``waterfall'' algorithm in ACL2 \cite{ACL2} both attempt to emulate
human reasoning processes. Compared to these studies, we place a
bigger emphasis on search, in order to be tolerant to mistaken steps,
and to try different heuristics in parallel. We also focus more on
heuristics for applying single theorems, although the system is
designed with the possibility of higher-level heuristics in mind (in
particular with the use of item types).

Finally, tactic-based automation such as \texttt{auto}, \texttt{simp},
and \texttt{fast} in Isabelle also use heuristics in the sense that
they apply theorems directionally, and are able to carry out
procedures. The main difference with our approach is the search
mechanism used. In tactic-based automation, the search is conducted
over the space of proof states, which consists of the current goal and
a list of subgoals. For \texttt{blast} and other tableau-based
methods, the search is over the space of possible tableaux. In our
approach, the search is saturation-based, and performed over the space
of propositions derivable from the initial assumptions.

A similar ``blackboard'' approach is used for heuristic theorem
proving by Avigad et al. \cite{avigad}, where the focus is on proving
real inequalities. The portion of our system concerning inequalities
is not as sophisticated as what is implemented there. Instead, our
work can be viewed as applying a similar technique to all forms of
reasoning.

\section{Conclusion}\label{sec:conclusion}

In this paper, we described an approach to automation in interactive
theorem proving that can be viewed as a mix of the currently
prevailing approaches. While the ideas behind the prover are mostly
straightforward, we believe the combination of these ideas is
underexplored and, given the examples above, holds significant promise
that warrants further exploration.

There are many aspects of \texttt{auto2} that can be improved in the
future. Two immediate points are performance and debugging. The
E-matching process is far from optimized, in the sense of
\cite{ematching}. For debugging, the program currently outputs the
list of updates applied to the state. One might instead want to view
and traverse the dependency graph of updates. One would also like to
query the rewrite table at any point in the proof.

There are also many directions of future research. I will just list
three main points:
\begin{itemize}
\item The scoring function is currently very simple. Except for a few
  cases, there is currently no attempt at take into account during
  scoring the proof step used. Instead, one would like to distinguish
  between proof steps that ``clearly should be applied'', and those
  that should be applied ``with reluctance''. There is also the
  possibility of using various machine learning techniques to
  automatically adjust the scoring function for individual proof
  steps.

\item Several aspects of elementary reasoning, such as dealing with
  associative-commutative functions, and with ordered rings and
  fields, pose special challenges for computers. While the current
  implementation is sufficient in these aspects for the examples at
  hand, more will need to be done to improve in both completeness and
  efficiency.

\item Finally, one would like to improve \texttt{auto2}'s ability to
  reason in other, diverse areas of mathematics and computer
  science. On the verification of imperative programs front, one would
  like to know how well \texttt{auto2} can work with separation logic,
  or perhaps a framework based on a mix of separation logic and
  ``natural'' reasoning used in the given examples will be ideal. On
  the mathematical front, each field of mathematics offers a
  distinctive system of heuristics and language features. One would
  like to expand the collection of proof steps, as well as proof
  script syntax, to reflect these features.
\end{itemize}

\paragraph{Acknowledgements.}
The author thanks Jasmin Blanchette for extensive comments on a first
draft of this paper, and also Jeremy Avigad, Adam Chlipala, and Larry
Paulson for feedbacks and discussions during the project. The author
also thanks the anonymous referees for their comments. The work is
done while the author is supported by NSF Award No. 1400713.

\end{document}